\documentclass[aps,prb,amsmath,amssymb,amsfont,showpacs,superscriptaddress,reprint]{revtex4-1}
\usepackage{amsmath}
\usepackage{xcolor}
\usepackage{graphicx}
\usepackage{braket}
\usepackage[artemisia]{textgreek}

\begin{document}

\title{Anisotropic Elliott-Yafet Theory and Application to KC$_8$ Potassium Intercalated Graphite}

\author{Bence G. M\'{a}rkus}
\affiliation{Department of Physics, Budapest University of Technology and Economics and MTA-BME Lend\"{u}let Spintronics Research Group (PROSPIN), P.O. Box 91, H-1521 Budapest, Hungary}

\author{L\'{e}n\'{a}rd Szolnoki}
\affiliation{Department of Physics, Budapest University of Technology and Economics and MTA-BME Lend\"{u}let Spintronics Research Group (PROSPIN), P.O. Box 91, H-1521 Budapest, Hungary}

\author{D\'{a}vid Iv\'{a}n}
\affiliation{Department of Physics, Budapest University of Technology and Economics and MTA-BME Lend\"{u}let Spintronics Research Group (PROSPIN), P.O. Box 91, H-1521 Budapest, Hungary}

\author{Bal\'{a}zs D\'{o}ra}
\affiliation{Department of Theoretical Physics, Budapest University of Technology and Economics and MTA-BME Lend\"{u}let Exotic Quantum Phases Group (Momentum), P.O. Box 91, H-1521 Budapest, Hungary}

\author{P\'{e}ter Szirmai}
\affiliation{Institute of Physics of Complex Matter, FBS Swiss Federal Institute of Technology (EPFL), CH-1015 Lausanne, Switzerland}

\author{B\'alint N\'afr\'adi}
\affiliation{Institute of Physics of Complex Matter, FBS Swiss Federal Institute of Technology (EPFL), CH-1015 Lausanne, Switzerland}

\author{L\'{a}szl\'{o} Forr\'{o}}
\affiliation{Institute of Physics of Complex Matter, FBS Swiss Federal Institute of Technology (EPFL), CH-1015 Lausanne, Switzerland}

\author{Ferenc Simon}
\affiliation{Department of Physics, Budapest University of Technology and Economics and MTA-BME Lend\"{u}let Spintronics Research Group (PROSPIN), P.O. Box 91, H-1521 Budapest, Hungary}

\keywords{graphite, intercalation compound, ESR, anisotropy, spin relaxation, Elliott-Yafet theory.}

\begin{abstract}We report Electron Spin Resonance (ESR) measurements on stage-I potassium intercalated graphite (KC$_8$). Angular dependent measurements show that the spin-lattice relaxation time is longer when the magnetic field is perpendicular to the graphene layer as compared to when the magnetic field is in the plane. This anisotropy is analyzed in the framework of the Elliott-Yafet theory of spin-relaxation in metals. The analysis considers an anisotropic spin-orbit Hamiltonian and the first order perturbative treatment of Elliott is reproduced for this model Hamiltonian. The result provides an experimental input for the first-principles theories of spin-orbit interaction in layered carbon and thus to a better understanding of spin-relaxation phenomena in graphene and in other layered materials as well.\end{abstract}

\maketitle

\section{Introduction}

Two dimensional layered materials are in the forefront of interest since the discovery of graphene \cite{novoselov2004}. These materials are atomically thin layers, i.e. they represent the ultimate limit for any circuit element including an electrode or a field effect transistor, etc. The low dimensionality is accompanied by rich novel phenomena, including massless Dirac fermionic behavior, robust quantum Hall effect, huge carrier mobility \cite{Neto2009}, and many more.

Among the compelling properties, the applicability of graphene for spintronics purposes attracted significant attention. Spintronics \cite{Zutic2004} intends to replace conventional electronics to yield a faster and more economic informatics architecture. The utility of spintronics in any material relies on the knowledge and theoretical description of the spin-relaxation time, $\tau_{\text{s}}$, i.e. the characteristic decay time of a non-equilibrium spin population. The initial reports on $\tau_{\text{s}}$ in graphene were conflicting concerning both its value and the dominant mechanism of spin-relaxation \cite{TombrosNAT2007,KawakamiBilayer,GuntherodtBilayer}. It became clear recently that the experimental data could be best described by the presence of extrinsic impurities (such as e.g. covalently bound H), which results in a significant contribution to the spin-relaxation \cite{Fabian2015,Nafradi2015}.

Alkali atom intercalated graphite re-emerged as a model system of graphene: e.g. the stage-I LiC$_6$ or AC$_8$ (A $=$ K, Rb, or Cs) is a model system of strongly chemically doped mono-layer graphene \cite{Gruneis2009a,Gruneis2009b,VallaPRL2011,Chacon2013,Chacon2014} as the presence of alkali atoms decouples the graphene sheets in graphite and also rearranges the stacking from the conventional \emph{AB} (or Bernal) stacking to \emph{AA} \cite{Dresselhaus1981}. It was shown from a temperature dependent electron spin resonance study for stage-I graphite \cite{Fabian2012} that the spin-relaxation time can be explained by the conventional Elliott-Yafet theory of spin-relaxation of metals with inversion symmetry.

An additional observation of the ESR studies was a moderate anisotropy of the ESR linewidth which corresponds to an anisotropic spin relaxation \cite{Fabian2012,Muller1962,Poitrenaud1970,Lauginie1980}. While it is not unexpected given the layered structure of intercalated graphite, we are not aware of a quantitative neither a qualitative description. Here, we report a detailed angular dependent ESR study of the linewidth anisotropy in KC$_8$. The result confirms the earlier indications and provides robust data. We present a model spin-orbit Hamiltonian to explain the observation and the perturbative treatment of Elliott \cite{Elliott1954} is reproduced for the case of anisotropy.

\section{Experimental}

The potassium intercalated stage-I graphite compound, KC$_8$ was prepared from Grade SPI-1 HOPG disc (SPI Supplies) with a diameter of $3$ mm and the thickness of $50-70$ {\textmu}m. The HOPG was annealed at $400^{\circ}$C under high vacuum before the intercalation to remove any remaining contamination. The sample was prepared under an Ar filled glove box to avoid oxygen and water exposure. The intercalation was performed using the two-zone vapor phase method \cite{Rudorff1954,Herold1955,Croft1960,Dresselhaus1981}, for this a special quartz tube was used, sealed under low pressure He. The temperature was held at $250^{\circ}$C with a gradient of $5^{\circ}$C during the 2-day long process. Stage-I stoichiometry was identified by the golden-yellow color of the samples and by ESR line shape and linewidth \cite{Muller1962,Poitrenaud1970,Lauginie1980,Fabian2012}. A photograph of final product can be seen in the inset of Fig. \ref{fig:spectra}. ESR measurements were carried out on a commercial Bruker Elexsys E500 X-band spectrometer at room temperature.

\section{Electron Spin Resonance spectroscopy results}

\begin{figure}[h!]
\includegraphics*[width=0.65\linewidth]{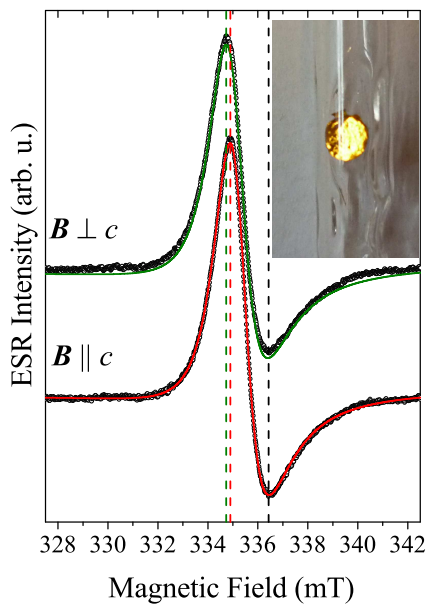}
\caption{ESR spectrum of the KC$_8$ sample with magnetic field parallel (bottom) and perpendicular (top) to the $c$ crystal axis. Experimental data are presented with open circles, solid lines are Dysonian fits. Dashed lines are guides to the eye to emphasize the anisotropy present in the width of the two lines. Inset shows a photograph of the material sealed in a quartz capillary.}
\label{fig:spectra}
\end{figure}

The room temperature ESR spectra of KC$_8$ HOPG are presented in Fig. \ref{fig:spectra} with a photograph of the sample sealed in the quartz tube. The bottom and top spectra are recorded in magnetic field directions parallel and perpendicular to the c axis, respectively. In both cases an asymmetric line was observed, which is identified as a Dysonian \cite{Feher1955,Dyson1955}. This line shape is known to appear in metallic materials, where the skin depth ($\delta$) is significantly smaller than the sample size ($d$). The solid lines of Fig. \ref{fig:spectra} are the fitted Dysonian lines.

The fitted parameters indicate that in our case the so-called "NMR-limit" is realized, when the diffusion time ($T_{\text{D}}$) is greater than the spin relaxation time ($\tau_{\text{s}}$). This phenomenon is understood and described by Walmsley and co-workers \cite{Walmsley1989} in $n$-doped graphite intercalation compounds (GIC). In this limit, the Dysonian line is the sum of an absorptive and dispersive Lorentzian curves \cite{slichter,Walmsley1996}.

The mean value of the linewidth is $1.28$ mT, which is in a good agreement with the literature results: $1.43$ \cite{Muller1962,Poitrenaud1970}, $1.14$ mT for $B \parallel c$ and $1.26$ mT for $B \perp c$ \cite{Lauginie1980}, $1.2$ mT for $B \parallel c$ and $1.3$ mT for $B \perp c$ \cite{Fabian2012}. This agreement confirms that the sample is indeed the stage-I KC$_8$ in agreement with the visual identification. An important observation is that the Dysonian linewidths differ for the two orientations by $0.09$ mT, which is denoted with dashed vertical lines in Fig.~\ref{fig:spectra}.

\begin{figure}[h!]
\includegraphics*[width=0.65\linewidth]{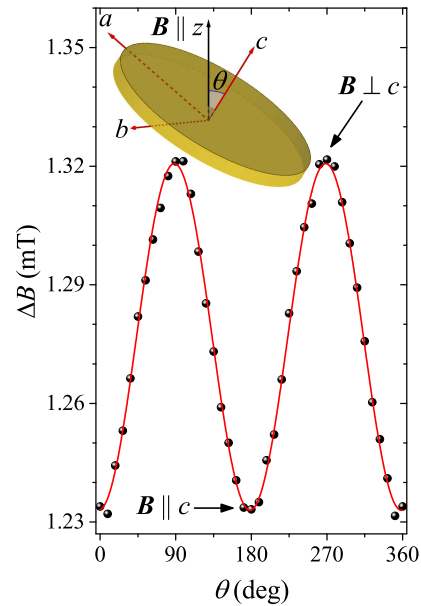}
\caption{Anisotropy of the ESR linewidth, when the sample is rotated against the magnetic field. The rotation angle, $\theta$ is measured between the $z$ axis, where the magnetic field points and the $c$ crystalline axis as shown in the inset. {\color{black}The error of the linewidth data is smaller than the symbols, however some systematic error arises from the uncertainty of the rotation angle.}}
\label{fig:rot}
\end{figure}

In Fig. \ref{fig:rot} the linewidth is plotted against the $\theta$ angle between the $c$ crystalline axis and the magnetic field, as shown in the inset of the figure. The width of the Dysonian in the two main directions are: $1.23$ mT for $\mathbf{B} \parallel c$ and $1.32$ mT for $\mathbf{B} \perp c$. The change in the width is continuous during the rotation. The anisotropy is thus $\mathrm{\Delta} B_{\perp}-\mathrm{\Delta} B_{\parallel} = 0.09$ mT, which is approximately $7\,\%$ of the mean value. It will be shown later, that this anisotropy is coming from the graphite host crystal.

{\color{black}We mention that the $g$-factor shift, $\mathrm{\Delta} g = g - g_{\text{e}}$, where $g_{\text{e}}$ is the free electron $g$-factor, also exhibit an anisotropy of $\mathrm{\Delta} g_{\perp}-\mathrm{\Delta} g_{\parallel} = 8.6(4) \times 10^{-4}$}, which is comparable to the shift itself. Unfortunately, the fact that $\mathrm{\Delta} g$ can only be measured with a low precision and requires a reference material directs the attention to the $\mathrm{\Delta} B$ linewidth, which does not suffer from these problems.

\section{Theoretical model to describe anisotropic spin relaxation}

Elliott \cite{Elliott1954} and Yafet \cite{Yafet1963} showed that spin relaxation in metals and semiconductors with inversion symmetry is caused by spin-orbit interaction (SOC). In the theory the conduction band $\left(\ket{1}\right)$ and a near lying band $\left(\ket{2}\right)$ is taken into account and the SOC is treated as a first order of perturbation.

The usual form of the isotropic SO coupling reads:
\begin{equation}
H_\text{SOC}=\lambda \mathbf{l \cdot s},
\end{equation}
where $\lambda$ is the spin-orbit coupling constant, $\mathbf{l}$ and $\mathbf{s}$ are the angular momentum and spin vector operators, respectively. For the case of an anisotropic material the effective Hamiltonian can be generalized as a bilinear form of $\mathbf{l}$ and $\mathbf{s}$:
\begin{equation}
H_\text{SOC}=\mathbf{l}\hat{\mathbf\Lambda}\mathbf{s},
\label{eq:aniso}
\end{equation}
where $\hat{\mathbf\Lambda}$ is a symmetric, positive definite matrix, called the SOC tensor. In the coordinate system, where this tensor takes a diagonal form, the Hamiltonian simplifies to
\begin{equation}\label{eq:diagSOC}
H_\text{SOC} = \lambda_x l_x s_x + \lambda_y l_y s_y + \lambda_z l_z s_z.
\end{equation}

Even though there are two degenerate spin states in the conduction band, Yafet showed that the SOC does not lift the degeneracy due to time reversal and the inversion symmetry (also known as Kramers degeneracy theorem \cite{Kramers1930}). Therefore a regular first order perturbation calculation can be applied on these states as well:
\begin{equation}
	\ket{\widetilde{1,\sigma}} = \ket{1,\sigma} +
			\sum_{\sigma'} \frac{\braket{ 2,\sigma' | H_\text{SOC} | 1,\sigma}}{\Delta} \ket{2,\sigma'},
\end{equation}
where $\ket{1}$ and $\ket{2}$ are the conduction and a near lying band, respectively, the spin is denoted with $\sigma$, $\Delta$ is the band separation, and $\ket{\widetilde{\,\quad}}$ denotes the perturbed states. Substituting Eq.~\eqref{eq:diagSOC} and applying it to the spin up state:
\begin{align}
	\ket{\widetilde{1,\uparrow}} = &\ket{1,\uparrow} + \frac{1}{2\Delta} \left[ \lambda_z \braket{2|l_z|1} \ket{2,\uparrow} + \right. \nonumber \\
			&\left. \left(	\lambda_x \braket{2|l_x|1} + \mathrm{i} \lambda_y \braket{2|l_y|1} \right) \ket{2,\downarrow}\right].
\end{align}

The $g$-factor shift can be calculated through the energy split caused by the Zeeman Hamiltonian. Assuming that the $B$ magnetic field is parallel to the $z$ axis this term has the following form:
\begin{equation}
	H_\text{Z}=\mu_\text{B}B(l_z+g_\text{e}s_z),
\end{equation}
where $g_\text{e}=2.0023$ is the free electron $g$-factor, $\mu_\text{B}$ is the Bohr magneton.

Degenerate perturbation theory can be applied to the conduction band, however, it turns out that the off-diagonal matrix elements of $H_\text{Z}$ are $0$ in the first order. The $g$-factor shift is caused by the nonzero expectation value of $l_z$ in the perturbed states, since the expectation value of $s_z$ only changes in second order, from $\pm 1/2$. Due to symmetry:
\begin{equation}
\braket{\widetilde{1,\uparrow} | l_z | \widetilde{1, \uparrow }} = - \braket{\widetilde{1,\downarrow}| l_z |\widetilde{1, \downarrow}}.
\end{equation}
Thus, the $g$-factor shift can be expressed as:
\begin{equation}
\mathrm{\Delta} g = 2 \braket{ \widetilde{1,\uparrow } | l_z | \widetilde{1, \uparrow}} = \frac{2 \lambda_z}{\Delta} \left| \braket{1|l_z|2} \right|^2.
	\label{eq:deltag}
\end{equation}

For the case of KC$_8$ a diagonal $\hat{\mathbf \Lambda}$ can be assumed, where the matrix elements, that connect the in-plane angular momentum and the spins, are the same. Depending on whether one is interested in the $g$-factor shift for the in-plane or out-of-plane external magnetic field, the $z$ direction of the coordinate system can be chosen accordingly, denoted with $z$ and $z'$. The $H_\text{SOC}$ can be separated to an isotropic ($\lambda_\text{iso}$) and anisotropic ($\lambda_\text{anis}$) part.

\begin{align}
	H_{\text{SOC},\parallel} &= \left(\lambda_\text{iso} + \lambda_\text{anis}\right)l_z s_z +
		\lambda_\text{iso} \left( l_x s_x + l_y s_y \right) \\
	H_{\text{SOC},\perp}      &= \lambda_\text{iso} l_z s_z + 
					\left[ 
						\left( \lambda_\text{iso} + \lambda_\text{anis} \right) l_x s_x +
						\lambda_\text{iso} l_y s_y
					\right].
\end{align}

In both cases, the $g$-factor shift can be calculated by Eq.~\eqref{eq:deltag}:
\begin{align}
	\mathrm{\Delta} g_{\parallel} &= \frac{2\left( \lambda_\text{iso}+\lambda_\text{anis} \right)}{\Delta}\left| \braket{1|l_z|2} \right|^2, \\
	\mathrm{\Delta} g_{\perp} &= \frac{2\lambda_\text{iso}}{\Delta}\left| \braket{1|l_{z'}|2} \right|^2,
\end{align}
from where it can be seen that:
\begin{equation}
\mathrm{\Delta} g_{\perp} - \mathrm{\Delta} g_{\parallel} = - \frac{2 \lambda_\text{anis}}{\Delta}. 
\end{equation}
This means that the experimental value of $-2 \lambda_\text{anis}/\Delta = 8.6 \times 10^{-4}$, which can serve as input value for the calculation of the spin orbit coupling.

To calculate the spin relaxation, the electron-phonon interaction has to be taken into account. For this, an interaction term $H_\text{int}$ is assumed. The exact form of $H_\text{int}$ is not required. This term is taken into account as a time dependent perturbation. Following Elliott's calculations the momentum and the spin relaxation time is to be compared. The momentum relaxation is: {\color{black}
\begin{equation}
\frac{1}{\tau} \propto \left| \braket{ \widetilde{2,\uparrow} | H_\text{int} | \widetilde{1,\uparrow} }\right|^2 +\left| \braket{ \widetilde{2,\downarrow} | H_\text{int} | \widetilde{1,\uparrow} }\right|^2,
\end{equation}
where the second term describes spin-flipping which is much smaller than the usual spin conserving momentum scattering, it can thus be neglected.} For the spin relaxation time, the following relation holds:
\begin{equation}
\frac{1}{\tau_{\text{s}}} \propto \left| \braket{ \widetilde{2,\downarrow} | H_\text{int} | \widetilde{1,\uparrow} }\right|^2.
\end{equation}

After reproducing Elliott's calculations, in first order, the ratio of the two relaxation times for the two directions read:
\begin{align}
\left.\frac{\tau}{\tau_{\text{s}}}\right|_{B \parallel c} &\propto \frac{\left| \braket{ \widetilde{2,\downarrow} | H_{\text{int}} | \widetilde{1,\uparrow} }\right|^2 }{ \left| \braket{ \widetilde{2,\uparrow} | H_\text{int} | \widetilde{1,\uparrow}} \right|^2} \propto \left(\frac{\lambda_{\text{iso}}+\lambda_{\text{anis}}}{\Delta}\right)^2, \\
\left.\frac{\tau}{\tau_{\text{s}}}\right|_{B \perp c} &\propto \frac{\left| \braket{ \widetilde{2,\downarrow} | H_\text{int} | \widetilde{1,\uparrow} }\right|^2}{\left| \braket{ \widetilde{2,\uparrow} | H_\text{int} | \widetilde{1,\uparrow}} \right|^2} \propto \left(\frac{\lambda_{\text{iso}}}{\Delta}\right)^2.
\end{align}
Taking into account, that $1 / \tau_{\text{s}} \propto \mathrm{\Delta} B$, the anisotropy of the ESR linewidth is:
\begin{equation}\label{eq:deltabk}
\mathrm{\Delta} B_{\perp} - \mathrm{\Delta} B_{\parallel} \propto - \frac{2 \lambda_{\text{anis}} \lambda_{\text{iso}} + \lambda_{\text{anis}}^2}{\Delta^2}.
\end{equation}
Eq. \eqref{eq:deltabk} yields that the small anisotropic part of the spin-orbit coupling is enhanced by the isotropic part, which dominates the contribution.

As a result, the above assumption of an anisotropic SOC is capable of reproducing the experimental observation of an anisotropic ESR linewidth.

\section{Conclusions}

We showed that stage-I potassium doped graphite exhibits anisotropic ESR linewidth, thus, the spin relaxation time is different along the $c$ axis and in the $ab$ plane. A model calculation is presented to explain this result and to extend the conventional Elliott-Yafet theory to anisotropic materials.

\section{Acknowledgement}
B. D. is supported by the Hungarian Scientific Research Fund No. K101244. P. Sz., B. N. and L. F. acknowledge the support of the Swiss National Science Foundation (Grant No. 200021\_144419) and ERC advanced grant ''PICOPROP'' (Grant No. 670918).

\bibliography{kc8}

\end{document}